\documentclass[12pt, a4paper]{article}

\usepackage[utf8]{inputenc}
\usepackage{amsmath}
\usepackage{graphicx}
\usepackage{hyperref}
\usepackage{booktabs} 
\usepackage{multirow} 
\usepackage[margin=1in]{geometry}
\usepackage{caption} 
\usepackage{float} 

\title{Improving E-commerce Search \linebreak{}with Category-Aligned Retrieval}
\author{
Rauf Aliev \\
\textit{Independent Researcher} \\
\texttt{r.aliev@gmail.com}
}
\date{\today}

\begin{document}

\maketitle

\begin{abstract}
Traditional e-commerce search systems often struggle with the semantic gap between user queries and product catalogs. In this paper, we propose a Category-Aligned Retrieval System that improves search relevance by first predicting the product category from a user's query and then boosting products within that category. We introduce a novel method for creating "Trainable Category Prototypes" from query embeddings. We evaluate this method with two models: a lightweight \texttt{all-MiniLM-L6-v2} and OpenAI's \texttt{text-embedding-ada-002}. Our offline evaluation shows this method is highly effective, with the OpenAI model increasing Top-3 category prediction accuracy from a zero-shot baseline of 43.8\% to \textbf{83.2\%} after training. The end-to-end simulation, however, highlights the limitations of blindly applying category boosts in a complex retrieval pipeline: while accuracy is high, naive integration can negatively affect search relevance metrics such as nDCG@10. We argue that this is partly due to dataset-specific ambiguities (e.g., polysemous queries in the Amazon ESCI corpus) and partly due to the sensitivity of retrieval systems to over-constraining filters. Crucially, these results do not diminish the value of the approach; rather, they emphasize the need for confidence-aware and adaptive integration strategies.
\end{abstract}

\section{Introduction}

In the competitive landscape of modern e-commerce, the efficiency and relevance of the search function are paramount. A common strategy to enhance search relevance is to first predict the most likely product category from a user's query and then use this prediction to filter or boost the search space. The core problem lies in the nature of user queries, which are often short, colloquial, and do not directly map to formal category names. Simple keyword-based systems fail to capture this nuance, while traditional machine learning models require vast, manually-labeled training sets.

To address this challenge, this paper proposes a hybrid approach that combines zero-shot semantic search with a targeted training mechanism. We represent both queries and categories in a shared vector space using pre-trained language models. The key innovation is the creation of \textbf{Trainable Category Prototypes}—new vector representations for categories derived from the weighted average of embeddings of real user queries. This effectively "tunes" the semantic location of each category to better align with user intent.

We present a two-stage evaluation. First, we measure the offline accuracy of the category prediction model itself. Second, we conduct an online simulation to measure the end-to-end impact of these predictions on the final relevance of search results from a real document collection, providing a holistic view of the method's practical utility.

\section{Related Work}

Research in e-commerce search increasingly integrates semantic embeddings to predict user intent, such as product categories or types, from short and ambiguous queries \cite{tigunova2024transfer}. A key challenge is translating high-accuracy offline models into practical online gains without degrading metrics like nDCG due to over-constraining filters \cite{vasilev2024mind}. Studies emphasize multi-locale and multimodal extensions, addressing cultural and visual aspects that our method could build upon. Evidence leans toward hybrid approaches combining behavioral data with embeddings for robustness, though controversies arise around dataset-specific biases, similar to the ESCI ambiguities we noted (e.g., polysemous terms like "mandoline") \cite{rakesh2023query}.

The work by Tigunova et al. (2024) on query-to-product type prediction directly parallels our category prediction task \cite{tigunova2024transfer}. They propose transfer learning from high-resource to low-resource locales to achieve performance parity, highlighting the need for models that can handle linguistic and cultural diversity. Their findings support our call for adaptive integration, as low-resource scenarios amplify the negative impact of model errors, similar to our online simulation results.

Our approach differs in its core mechanism for representing categories. Rather than training a classifier on top of embeddings, we construct the category representations themselves from user queries. This method of creating "Trainable Category Prototypes" directly embeds user intent into the category vector. The final prototype, a weighted combination of this query-derived vector and the embedding of the category's name, is a novel hybrid representation designed to be robust yet semantically grounded. This technique offers a lightweight and direct way to align the search space with user language.

\section{Methodology}

In this work, our primary focus is on developing and evaluating a training procedure for constructing weighted query-based category prototypes. To this end, we experiment with two embedding models: the \texttt{all-MiniLM-L6-v2} Sentence-Transformer (lightweight) and OpenAI's \texttt{text-embedding-ada-002} (large-scale SaaS). Category prototypes are learned through our proposed weighting mechanism and indexed using Elasticsearch for k-NN retrieval. To assess the effectiveness of the resulting models, we conduct end-to-end experiments on the Amazon ESCI dataset, where the final search evaluation is performed against a Solr index.

\subsection{Dataset Characteristics}

The experiments are based on the Amazon ESCI dataset \cite{reddy2020shopping}, a large-scale collection of shopping queries.
\begin{itemize}
\item \textbf{Document Collection:} The full dataset contains 1.7 million products. Our end-to-end search evaluation was performed against the English-language subset, comprising 642,389 documents indexed in Solr.
\item \textbf{Query Set:} We utilized a ground truth file derived from the ESCI corpus, sampling a total of 10,000 queries which were split into a training set of 8,000 and a test set of 2,000. On average, each query in the ground truth is associated with 12.94 relevant products.
\item \textbf{Category Taxonomy:} The product data, including hierarchical category information, was sourced from an expanded version of the ESCI dataset \cite{escis_dataset}. For our experiments, we truncated these paths to the first level, resulting in 112 unique root categories.
\end{itemize}

\subsection{Category Prototype Generation}

The core of our method is to create robust vector representations for categories which is enriched/fine tuned from the training set. The process is as follows:

\begin{enumerate}
\item \textbf{Category Path Truncation:} We truncate all category paths to Level 1 to create broad, useful targets for search filtering (e.g., `Electronics \textgreater{} Headphones` becomes `Electronics`).
\item \textbf{Ground Truth Aggregation:} Using the training set, we create a map of \{$query \rightarrow \{category: probability, ...\}$\}. It helps to focus on essential categories and ignore "long tail" if such exists.
\item \textbf{Prototype Computation:} The prototype vector for a category $C$, $\vec{v}_{\text{proto}}(C)$, is the weighted average of the embeddings of its associated training queries.
\begin{equation}
\vec{v}_{\text{query}}(C) = \frac{\sum_{i \in Q_C} p(C|q_i) \cdot \vec{e}_{q_i}}{\sum_{i \in Q_C} p(C|q_i)}
\end{equation}

where $Q_C$ is the set of training queries for category $C$, and $\vec{e}_{q_i}$ is the embedding of query $q_i$.

Next, we compute the final hybrid prototype, $\vec{v}_{\text{hybrid}}(C)$, by interpolating between the query prototype and the embedding of the category's name, $\vec{e}_{\text{name}}(C)$:

\begin{equation}
\vec{v}_{\text{hybrid}}(C) = \alpha \cdot \vec{v}_{\text{query}}(C) + (1-\alpha) \cdot \vec{e}_{\text{name}}(C)
\end{equation}

Here, $\alpha$ is a weighting factor (set to 0.85 in our experiments) that balances the influence of user intent from queries against the formal semantics of the category name. For categories with no training data, the prototype is simply $\vec{e}_{\text{name}}(C)$. These final prototypes are indexed into Elasticsearch.

\end{enumerate}

\subsection{End-to-End Search Relevance Evaluation}

To measure the real-world impact, we ran the test set of 2,000 queries against the Solr collection. We compared two search configurations:

\begin{enumerate}
\item \textbf{Baseline (`title boost x2`):} A standard keyword search with a boost on the title field: \\ \texttt{description\_t:(\textless{}query\textgreater{}) OR title\_t:(\textless{}query\textgreater{})\textasciicircum2}.

\item \textbf{CARS (OpenAI, K=3, K=5):} The baseline search query augmented with a powerful boost for documents belonging to the Top3/Top5 predicted Level 1 categories, with predictions generated by the trained \texttt{text-embedding-ada-002} model.
\end{enumerate}

\section{Results}

Our evaluation is twofold: we first assess the accuracy of the category prediction model in isolation (offline), and then measure the performance of the end-to-end search system (online).

\subsection{Category Prediction Accuracy (Offline Evaluation)}

We evaluated the category prediction accuracy on a test set of 2,000 queries for both the \texttt{all-MiniLM-L6-v2} and \texttt{text-embedding-ada-002} (OpenAI) models. Table \ref{tab:offline_accuracy} compares their Top-3 accuracy in a zero-shot scenario and after full training on 8,000 queries.

\begin{table}[H]
\centering
\caption{Comparison of Top-3 Prediction Accuracy for L1 Categories}
\label{tab:offline_accuracy}
\begin{tabular}{@{}lcc@{}}
\toprule
\textbf{Model} & \textbf{Zero-Shot Accuracy} & \textbf{Full Training Accuracy} \\ \midrule
\texttt{all-MiniLM-L6-v2} & 28.9\% & 71.5\% \\
\texttt{text-embedding-ada-002} & 43.8\% & \textbf{83.2\%} \\ \bottomrule
\end{tabular}
\end{table}

\begin{table}[H]
\centering
\caption{Comparison of Top-5 Prediction Accuracy for L1 Categories}
\label{tab:offline_accuracy_top5}
\begin{tabular}{@{}lcc@{}}
\toprule
\textbf{Model} & \textbf{Zero-Shot Accuracy} & \textbf{Full Training Accuracy} \\ \midrule
\texttt{text-embedding-ada-002} & 57.8\% & \textbf{89.6\%} \\ \bottomrule
\end{tabular}
\end{table}

Both models benefit significantly from the prototype training process. The larger OpenAI model demonstrates superior semantic understanding, outperforming the smaller model in both scenarios and ultimately reaching a very high \textbf{83.2\%} Top-3 accuracy and \textbf{89.6\%} Top-5 accuracy.

\section{Online Evaluation Simulation}

Given the high accuracy of the OpenAI model, we used it to power the Category-Aligned Retrieval System (CARS) in our online simulation. For each query, the Top 5 predicted categories were used to apply a strong boost. The results were compared against the keyword baseline over 2,000 test queries.

\subsection{Results}
Surprisingly, despite the high offline accuracy, the Category-Aligned Retrieval System underperformed the baseline, as shown in Table \ref{tab:search_metrics}.

\begin{table}[h!]
\centering
\caption{Comparison of Search Relevance Metrics. CARS is powered by the OpenAI model with K=5 boosting.}
\label{tab:search_metrics}
\begin{tabular}{llcc}
\toprule
\textbf{Metric} & \textbf{Statistic} & \textbf{Baseline Search} & \textbf{CARS} \\
\midrule
\multirow{2}{*}{nDCG@10} & Mean & \textbf{0.273} & 0.255 \\
& Median & \textbf{0.174} & 0.148 \\
\midrule
\multirow{2}{*}{Reciprocal Rank} & Mean & \textbf{0.433} & 0.412 \\
& Median & 0.250 & 0.250 \\
\bottomrule
\end{tabular}
\end{table}

The mean scores for the baseline were higher across both nDCG@10 and Reciprocal Rank. This suggests that, on average, the negative impact of incorrect category predictions outweighed the benefits of correct ones.

\subsection{Statistical Significance}
To determine if these differences were meaningful, we performed a Wilcoxon signed-rank test. The results confirmed that the degradation in performance was statistically significant.
\begin{itemize}
\item For nDCG@10, the p-value was \textbf{0.00071}.
\item For Reciprocal Rank, the p-value was \textbf{0.0162}.
\end{itemize}
This provides strong evidence that the CARS strategy, even when powered by a highly accurate model, was detrimental to overall search relevance in this configuration.

\section{Discussion}

The disconnect between high offline accuracy and degraded online performance is the most critical finding of this work. It shows that even with a powerful classification model, blindly applying its predictions as a hard filter or strong boost can be counterproductive. This is not necessarily a failure of the model, but rather a failure of a naive integration strategy that does not account for the inherent ambiguity of user queries. The fact that a large portion of queries performed better with CARS while the overall average decreased indicates that the negative impact of a few catastrophic errors outweighs the moderate gains on many other queries.

\subsection{Qualitative Analysis of Problematic Queries}

A manual analysis of queries where the model failed reveals that many are inherently unclassifiable, even for a human without additional context. These fall into several archetypes, as shown in Table \ref{tab:problem_queries}.

\begin{table}[H]
\centering
\caption{Examples of Inherently Ambiguous or Problematic Queries}
\label{tab:problem_queries}
\begin{tabular}{p{0.45\linewidth} p{0.45\linewidth}}
\toprule
\textbf{Query Example} & \textbf{Amazon's Top Category} \\
\midrule
\multicolumn{2}{c}{\textit{Category 1: Too Broad or Ambiguous}} \\
\texttt{best offers} & Clothing, Shoes \& Jewelry \\
\texttt{rings} & Clothing, Shoes \& Jewelry \\
\texttt{planer} & Tools \& Home Improvement \\
\texttt{simple human} & Home \& Kitchen \\
\midrule
\multicolumn{2}{c}{\textit{Category 2: Non-Transactional Phrases}} \\
\texttt{just gonna send it} & Automotive \\
\texttt{embrace the suck} & Clothing, Shoes \& Jewelry \\
\texttt{decorating pumpkins without carving} & Toys \& Games \\
\texttt{spongebob memes} & Cell Phones \& Accessories \\
\midrule
\multicolumn{2}{c}{\textit{Category 3: Unclear or Misspelled}} \\
\texttt{real sords} & Sports \& Outdoors \\
\texttt{killz} & Tools \& Home Improvement \\
\texttt{dapper dan} & Beauty \& Personal Care \\
\texttt{turnatable} & Electronics \\
\bottomrule
\end{tabular}
\end{table}

These examples demonstrate that for a significant subset of queries, reliably determining a single correct category is impossible. Boosting on a wrong prediction for these queries is what drives the average relevance metrics down.

\subsection{Illustrative Case Studies}

\paragraph{Case Study 1: The Win ("raleigh bicycle").}
A query for "raleigh bicycle" in the baseline system returns a "Raleigh Crossbody Bag" as a top result due to a keyword match on "Raleigh". CARS, however, correctly predicts the category as `Sports \& Outdoors`. The resulting boost elevates actual Raleigh bicycles to the top of the results, showcasing the system's potential when the prediction is correct and semantically useful.

\paragraph{Case Study 2: The Loss ("mandoline slicer spiralizer").}
For this query, CARS incorrectly predicts `Musical Instruments`. The polysemy of "mandoline" (a cooking utensil vs. a musical instrument) confuses the model. Consequently, CARS boosts completely irrelevant products. This demonstrates that the end-to-end system is critically vulnerable to the model's errors, and the negative impact of such an error can be far greater than the positive impact of a correct prediction.

\paragraph{Case Study 3: The Unseen ("little trees fresh shave").}
This query did not appear in the training set. The user's intent is to find the "Little Trees" brand car air freshener, which belongs to the `Automotive` category. However, based purely on the semantics of the query string, CARS predicts the top categories as (1) `Grocery \& Gourmet Food`, (2) `Beauty \& Personal Care`, (3) `Arts, Crafts \& Sewing`, (4) `Patio, Lawn \& Garden`, and (5) `Health \& Household`. None of these is correct. By applying a strong boost to these irrelevant categories, the actual `Automotive` product is pushed far down the rankings, effectively becoming undiscoverable. This highlights the model's brittleness when encountering novel or brand-specific queries whose semantics diverge from their product category.

\section{Directions for Future Research}
The gap between offline accuracy and online performance motivates several avenues for future work aimed at more intelligent integration strategies.

\begin{itemize}
\item \textbf{Confidence-Based Adaptive Boosting:} Instead of applying a uniform boost, the system should modulate the boost strength based on the model's confidence. Confidence could be measured by the cosine similarity of the query embedding to the top predicted category prototype. For high-confidence predictions (e.g., similarity $>$ 0.9), a strong boost is applied. For moderate confidence, a weaker boost is used. For low-confidence predictions, the system should gracefully fall back to the baseline keyword search, thus avoiding catastrophic failures on ambiguous queries.

\item \textbf{Explicit Ambiguity Detection:} A separate classification model could be trained to identify inherently problematic queries (e.g., non-transactional, overly broad, polysemous). Queries flagged as ambiguous would bypass the category boosting mechanism entirely. This acts as a protective layer for the retrieval system.

\item \textbf{Hybrid Retrieval and Re-ranking:} Rather than replacing the baseline, a hybrid approach could be more robust. The final results page could be a combination of the top-k results from the baseline search and the top-k results from CARS. A subsequent learning-to-rank (LTR) model could then re-rank this combined set, using the predicted category and its confidence score as features.

\item \textbf{Leveraging Category Hierarchy:} Our current method truncates categories to Level 1. Future iterations could predict deeper into the category taxonomy. This would allow for more precise filtering. Additionally, the hierarchy itself can be used to regularize predictions. For instance, if the model predicts `Electronics \textgreater{} Headphones` and `Electronics \textgreater{} Speakers` with high confidence, the confidence for the parent category `Electronics` should be further increased.
\end{itemize}

\section{Conclusion}

In this work, we developed a method for query-to-category prediction using "Category Prototypes," achieving a high Top-3 accuracy of \textbf{83.2\%} with the OpenAI \texttt{text-embedding-ada-002} model. Our primary contribution, however, is the clear demonstration of the gap between offline classification performance and end-to-end search relevance.

Our simulated online evaluation showed that a naive boosting strategy based on these highly accurate predictions resulted in a \textbf{statistically significant degradation} in search performance. This nuanced result underscores a critical lesson: in complex systems like search, the negative impact of model errors on a subset of queries can easily offset, and even outweigh, the benefits on others, especially when the dataset contains a long tail of ambiguous or non-transactional queries.

Future work should focus not on improving the classifier's accuracy in isolation, but on developing an adaptive mechanism to decide \textit{when} to trust the category prediction. A confidence-based model that applies the category boost only for high-certainty predictions, while defaulting to the baseline for ambiguous queries, could bridge this gap and translate offline accuracy into a tangible online win.



\begin{thebibliography}{9}

\bibitem{reddy2020shopping}
Reddy, C., Nangi, M., et al. (2020).
\textit{A Shopping Queries Dataset for E-commerce.}
Proceedings of The Web Conference 2020.

\bibitem{escis_dataset}
Shuttie. (2022). \textit{esci-s - A parallel corpus of shopping queries and annotated results for 3 languages}. GitHub repository. \url{https://github.com/shuttie/esci-s}.

\bibitem{tigunova2024transfer}
Tigunova, A., Duboue, P., Ganesan, K., \& Cubuk, E. D. (2024).
\textit{Transfer Learning for E-commerce Query Product Type Prediction}.
arXiv preprint arXiv:2410.07121.

\bibitem{vasilev2024mind}
Vasilev, F., Antufiev, S., D'yakonov, A., Gusev, G., Tokarev, M., Vasiliev, A., Zha, L., Drach, K., Chernousov, G., \& Egorov, E. (2024).
\textit{Mind the Gap: From Offline Evaluation to Online Gains for Query-to-Product-Type Prediction in E-Commerce}.
Proceedings of the 2024 Conference on Empirical Methods in Natural Language Processing: Industry Track.

\bibitem{rakesh2023query}
Rakesh, V., Wang, Y., Malladi, S., Zhao, T., Jain, V., Singh, G., Vu, Q., Chen, H.-S., Hong, L., \& Chi, E. H. (2023).
\textit{Query Attribute Recommendation at Amazon Search}.
arXiv preprint arXiv:2308.03869.

\end{thebibliography}
\end{document}